\documentclass[aps,reprint,superscriptaddress,nofootinbib, notitlepage,prl]{revtex4-2}

\usepackage[utf8]{inputenc}
\usepackage{graphicx}
\usepackage{xcolor}
\usepackage{mathrsfs}
\usepackage{amsmath}

\begin{document}
\frenchspacing

\title{Does fluid structure encode predictions of glassy dynamics?}
\author{Tomilola M. Obadiya}\email{tomilola.medetin.obadiya@emory.edu}
\author{Daniel M. Sussman}\email{daniel.m.sussman@emory.edu}
\affiliation{Department of Physics, Emory University, Atlanta, GA, USA}

\begin{abstract}
Data-driven approaches that infer the local structures responsible for plasticity in amorphous materials have made substantial contributions to our understanding of the failure, flow, and rearrangement dynamics of supercooled fluids. Some of these methods, such as the ``softness'' approach, have identified combinations of local structural features in a supercooled particle's environment that predict energy barriers associated with particle rearrangements. This approach also predicts the onset temperature, often characterized as the temperature below which the system's dynamics becomes non-Arrhenius and above which local structures are no longer predictive of dynamical activity. We implement a transfer-learning approach in which we first show that classifiers can be trained to predict dynamical activity even far above the onset temperature. We then show that applying these classifiers to data from the supercooled phase recovers the same essential physical information about the relationship between local structures and energy barriers that softness does.
\end{abstract}

\maketitle

\section{I. Introduction}

At the microscopic scale glassy materials possess amorphous structures very similar to their dense fluid phase \cite{berthier2011dynamic}, but their dynamical properties are radically different \cite{angell1995formation}. Many different temperature scales have been identified in the transition from equilibrium fluid to out-of-equilibrium amorphous solid; the first one encountered upon cooling the system is the onset temperature, $T_0$, which marks a transition to ``landscape-influenced'' dynamics  \cite{sastry1998signatures,debenedetti2001supercooled,banerjee2017determination,folena2020rethinking}. Above this temperature the particle-level dynamics are simple, diffusive, and do not depend on local structure; below this temperature dynamics become both spatially and temporally heterogeneous, depend strongly on local structure, and (for fragile glassformers) begin slowing down super-exponentially \cite{keys2011excitations, ediger2000spatially, berthier2011dynamical, ediger1996supercooled,pedersen2018phase}. 

The role of local microscopic structure on local dynamics is a deep question, and a major theme of recent research has been a search for correlations between them \cite{berthier2007structure,kawasaki2007correlation, patrick2008direct,jack2014information,manning2011vibrational, widmer2008irreversible}. Data-driven approaches, using numerical simulations to generate large data sets for the training of Support Vector Machines (SVMs) \cite{cubuk2015identifying,schoenholz2016structural,sussman2017disconnecting}, Graph Neural Networks \cite{bapst2020unveiling}, or other techniques \cite{boattini2020autonomously,boattini2021averaging} have shown substantial promise in finding maximally correlative structures for the dynamics at different timescales in strong glassformers \cite{cubuk2020unifying}, fragile glassformers \cite{schoenholz2016structural,bapst2020unveiling,jung2022predicting}, and even anomalous glassformers modeled after biological systems \cite{tah2021quantifying}. Notably, approaches based on SVMs \cite{schoenholz2016structural} have found \emph{physically interpretable} classifiers. This physical interpretation comes from training classifiers at low temperatures and studying how they behave when applied to data at other temperatures: by maximizing correlations in the training data in a compressed way, these classifiers are learning combinations of structural features that can be interpreted as a local energy barrier to particle rearrangements. However, relatively little is known about why these approaches work, and whether or how one could use these results to help build a more robust theoretical description of glass-forming systems. Key unresolved questions include why these particular approaches lead to what is apparently a local order parameter for the supercooled liquids, and how the learned energy barriers actually depend on the construction of the classifiers.

In this work we bridge between the changing dynamical behavior above and below the onset temperature on the one hand and the physical interpretation of amorphous state classifiers on the other. We first demonstrate that the same machine learning techniques that have successfully correlated structure and dynamics in the supercooled phase can be used to classify ``extreme diffusive'' events even far above the onset temperature. In the spirit of a transfer learning approach, we show that these liquid-state classifiers can statistically identify activated events in the supercooled phase, even though the character of the activated dynamics below $T_0$ changes dramatically. We further show that not only can accuracy on a classification task be maintained, but that the physical interpretability is maintained: apparently fluid-phase classifiers also learn energy barriers in the super-cooled phase.

\section{II. Methods}
\subsection{Model and Simulations}
Our analysis is focused on a large set of molecular dynamics simulations of $N=4096$ particles, using the standard 80:20 Kob-Andersen model \cite{kob1995testing} (with a cutoff distance of $2.5$) at a density of $\rho=1.2$ in a cubic box with periodic boundary conditions. Throughout we report all quantities in dimensionless (reduced) units, using the standard Lennard-Jones (LJ) convention in which the base units are distance (measured in units of the large particle diameter, $\sigma_{AA}$), energy (in units of the interaction parameter, $\epsilon_{AA}$), and mass (in units of the particle mass, $m$). For this model the dimensionless onset temperature is often reported as $T_0\approx 0.87$ \cite{keys2011excitations,schoenholz2016structural}; given the broad crossover in the dynamics, values between $0.8$ and $1.2$ are also reasonable estimates for this temperature scale \cite{debenedetti2001supercooled,pedersen2018phase}. Our simulations were done in the NVT ensemble for temperatures in the range $T \in [0.45, 2.0]$ in standard LJ units (i.e., ranging from the moderately supercooled to the supercritical fluid phase). We note that our use of a deterministic Nose-Hoover thermostat \cite{frenkel2023understanding} should not unduly influence the results reported below: in our classification task we do not include any information about the fictitious degrees of freedom used by the thermostat to maintain an average temperature, and we expect that if anything our quantitative results would improve upon conducting simulations in an NVE ensemble.

All simulations  were conducted using HOOMD-Blue \cite{anderson2020hoomd}, and in the following we focus on the behavior of the large particles. Our simulation configurations were generated as follows. We first equilibrated a system with random initial conditions $5000\tau$ at a temperature of $T=0.45$. We used the final configuration of this as an initial seed for out other simulations: a snapshot was loaded as the initial configuration for our other simulations, each of which was allowed to equilibrate for 1000$\tau$ at its target temperature. After this, data used in this study was saved at intervals of $1\tau$.

We begin by characterizing the local structural environment of particles in these simulations, using two-point radial structure functions $G_X(i;r,\delta)$ \cite{behler2007generalized}. For a target particle $i$, these functions are defined as
\begin{equation} 
G_X (i;r,\delta) = \sum_{j \in X} \exp\left( \frac{-(r - R_{ij})^2}{2\delta^2}\right),
\end{equation}
where $X$ denotes which of the components of the binary mixture is being considered, $r$ is a parameter controlling the distance from which dominant contributions to the feature come, $\delta$ is a parameter controlling the width of the Gaussian shells, and $R_{ij}$ is the distance between particles $i$ and $j$. We characterize the local environment of particle $i$ with a vector in a 100-dimensional feature space, $\vec{F}_i$, with $\delta=0.2$, $0<r<5$ in increments of $0.1$, and $X=A,\ B$. Each feature is standardized \cite{hsu2003practical} so that at the training temperature they have zero mean and unit variance.

We next choose a  measure for the dynamics of particles; to be consistent with work on activated dynamics we use $p_{hop}$ as introduced in Ref.~\cite{candelier2009building}. We use an observational time window of $10$ LJ time units, for which
\begin{equation*}
p_{hop}(i,t) = \sqrt{\langle (\vec{r}_i(t) - \langle \vec{r}_i \rangle_{w_2})^2\rangle_{w_1} \langle (\vec{r}_i(t) - \langle \vec{r}_i \rangle_{w_1})^2\rangle_{w_2}},
\end{equation*}
where ${w_1} = [t-5, t]$, ${w_2} = [t,t+5]$, and thus $\langle \cdots \rangle_{w_i}$  averages over one half of the observation window. We do not believe that using $p_{hop}$ as a dynamical label is crucial -- preliminary results indicate that choosing instead to measure particle dynamics using their cumulative displacement over the same time window leads to qualitatively identical results -- and we note that length of time window is optimized for detecting activated events in the supercooled regime. 

\subsection{Machine learning protocol}

We train SVMs connecting structural features with dynamic observables largely following the ``softness'' methodology \cite{schoenholz2016structural}. We build a training set by combing through MD trajectories for examples of dynamically active (``rearranging'') and inactive (``non-rearranging'') particles, and train a linear soft margin SVM (using the Scikit-learn package \cite{pedregosa2011scikit}) to classify these examples. We can then use the learned classifier (here: a hyperplane in feature space) to try to predict dynamics based on a particle's instantaneous environment, and we define the softness of particle $i$ at time $t$, $S_i(t)$, as the shortest distance between its vector of structural features and this classifying hyperplane.

The training set construction for our ``softness'' classifier closely followed the protocol outlined in Ref.~\cite{schoenholz2016structural}. We constructed a balanced $7600$-sample training set using the coldest temperature considered ($T = 0.45$): $3800$ rearranging samples and $3800$ non-rearranging samples. We adopted the previously-used convention of associating the structural data of a particle $i$ at time $t - 2\tau$ with the dynamical state at time $t$. We defined a rearranging particle if, at time $t$,  $p_{hop}(i,t) > p_{c}$ where $p_c =0.2$. We defined a non-rearranging particle by requiring its $p_{hop}(i,t)$ value to remain less than a lower threshold of $p_{l}=0.0085$ for at least $120\tau$ duration of time. We then used the local structure of the non-rearranging particle in the middle of its time of low activity. Unlike in previous work, we take structure and $p_{hop}$ values directly from the thermal configurations rather than quenching to the inherent states (in part because the fluid-phase simulations would be far from any minima). Unless otherwise stated, we used a soft-margin misclassification hyperparameter of $C=10^{-2}$.

A major finding of Ref.~\cite{schoenholz2016structural} was that this signed distance -- softness -- encodes the probability the target particle would rearrange at a given temperature. The corresponding curves for the probability of rearranging at different values of $S$ as a function of $T$ all intersected at a common temperature, which in turn suggested the existence of an onset temperature above which structure was no longer predictive of dynamical events. The predicted value of $T_0$ was consistent with alternative definitions \cite{debenedetti2001supercooled,keys2011excitations,schoenholz2016structural} and with the numerical values cited above. Before we return to this finding, we first ask: Can we learn to classify dynamical events based on structure not in the supercooled regime but at and even \emph{above} the onset temperature?

For $T>T_0$ individual particle motion is diffusive rather than activated, and it is not clear that using $p_{hop}$ as a dynamical label is the most natural choice. We continue to use it as an indicator function -- it is still large for dynamical trajectories that move a particle far from its initial position and small for diffusive motions that stay near a particle's initial position -- and will show in a later work that this choice is not crucial to our results. To identify ``extreme events'' to classify, we select particles in high and low tails of the probability density function of $p_{hop}$ at different temperatures. To have similarly sized training sets as in the case of softness, we choose lower and upper cutoffs ($p_l$ and $p_u$) that captured the most extreme $0.033\%$ of low- and high-activity events, respectively. We identify particle $i$ at a given time $t$ as ``extremely diffusive'' if $p_{hop}(i,t) > p_u$, and associate it with the particle's local structure at time $t - 2\tau$. Similarly, if  $p_{hop}(i,t) < p_l$, the particle is  identified as  ``extremely non-diffusive,'' and its structure at time $t - 2\tau$ is included in the training set. The training set for each temperature we considered above $T_0$ contained $10400$ balanced samples. 

Aside from this difference in choosing ``rearranging'' and ``non-rearranging'' labels, we follow the methodology above: we find a linear soft margin SVM that best classifies a labeled training set, and then apply this classifier to new data. To distinguish it from softness we call the distance of a point in feature space to such a classifying hyperplane the ``fluidity,'' and we use we $\mathcal{F}^{T}_i$ to denote the fluidity of particle $i$ with respect to a classifier trained from data at temperature $T$. Given any of our classifiers, one can compute a particle's softness or fluidity by computing its feature vector (which depends only on the instantaneous structure around the particle) and evaluating $\alpha_i(t) = \vec{w}_\alpha\cdot \vec{F}_i(t) -b_\alpha,$ where $\vec{w}_\alpha$ is the normal vector and $b_\alpha$ the bias defining a classifying hyperplane, and where $\alpha$ refers to either softness $S$ or a fluidity $\mathcal{F}^T$. We note that the term ``fluidity'' has previously been used to describe an average rate of plastic events in models of soft glassy rheology  \cite{bocquet2009kinetic}; while our definition is different, we will see that highly ``fluid'' particles have more active dynamics at high temperatures and, indeed, are more likely to undergo plastic rearrangement events at low temperatures.

\section{III. Classification in the fluid phase}

\begin{figure}[ht]
    \centering
    \includegraphics[width =1.0\linewidth]{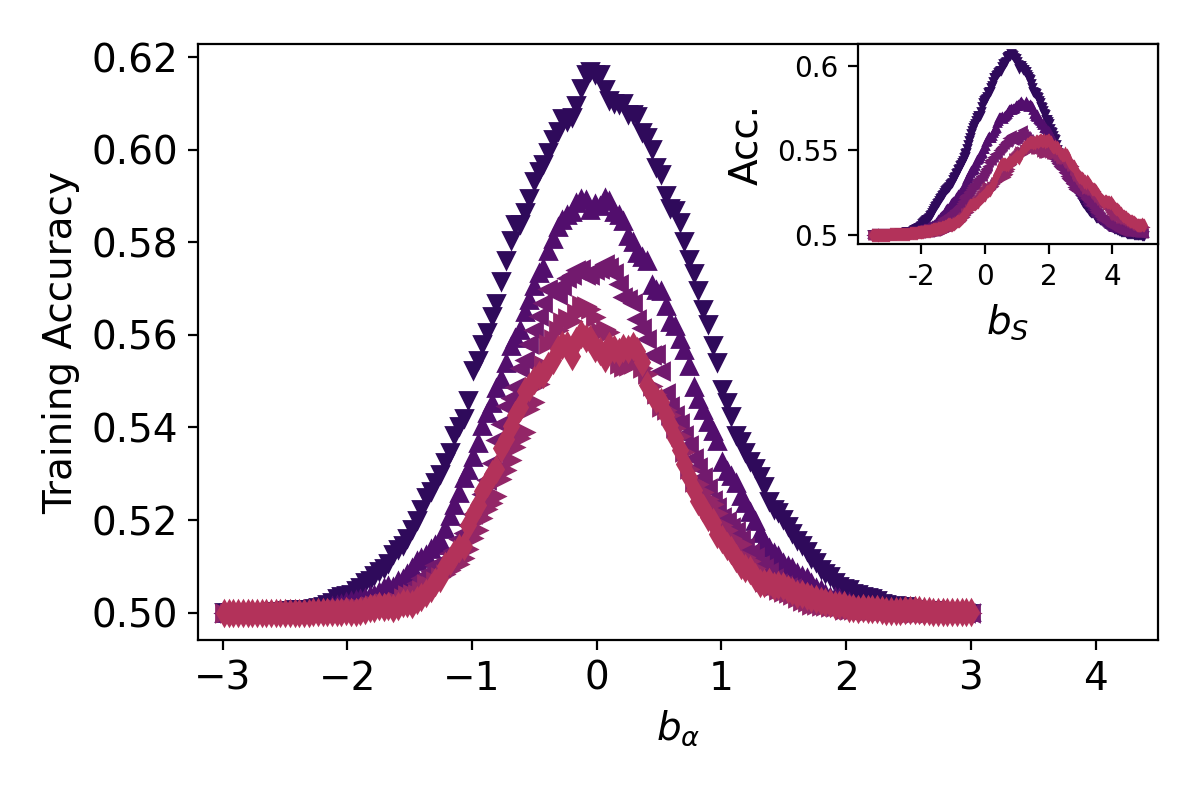}
    \caption{\textbf{Fluidity classifies rare events at high and low temperatures} The points show the 5-fold cross-validation accuracy  of linear SVMs trained on extreme diffusive samples at $T=1.0,1.2,1.4,1.8,2.0$ (dark blue to light red) as a function of the classifier's bias. Each classifier achieves near-peak accuracy for small values of the bias. In contrast, the inset show the test classification accuracy of the ``softness'' classifier trained on activated dynamics at T=0.45 and applied to the extremes of diffusive events different temperatures, for which very different values of the bias optimize performance.}
    \label{fig:classificationAccuracy}
\end{figure}

We find that we \emph{can} learn to classify extreme diffusive events even far above the onset temperature using local structure. Hyperplanes are characterized by a normal vector and a bias; the direction of the normal corresponds to the linear combination of features that has been learned, and the bias is an offset that bests separates the training set given that direction. We expect the \emph{direction} to encode the key physical features governing rearrangements, whereas we expect the \emph{bias} may be strongly dependent on details such as the choice of time window or the temperature of the training set. For instance: with our fixed-threshold definition of a rearrangement the total number of rearranging particles increases as $T$ increases, so even if the same underlying structural variable controls rearrangements the optimal bias of the hyperplane will shift to maximize the soft margin in the training set data. Because of this, we want to remove the influence of the bias on our later results. In Fig.~\ref{fig:classificationAccuracy} we show the training accuracy of fluidity as a function of the bias, and during our transfer learning approach later we will select values that maximize our classification accuracy not on the training but on a low-temperature test set.

\begin{figure}[t!]
    \centering
    \includegraphics[width =1.\linewidth]{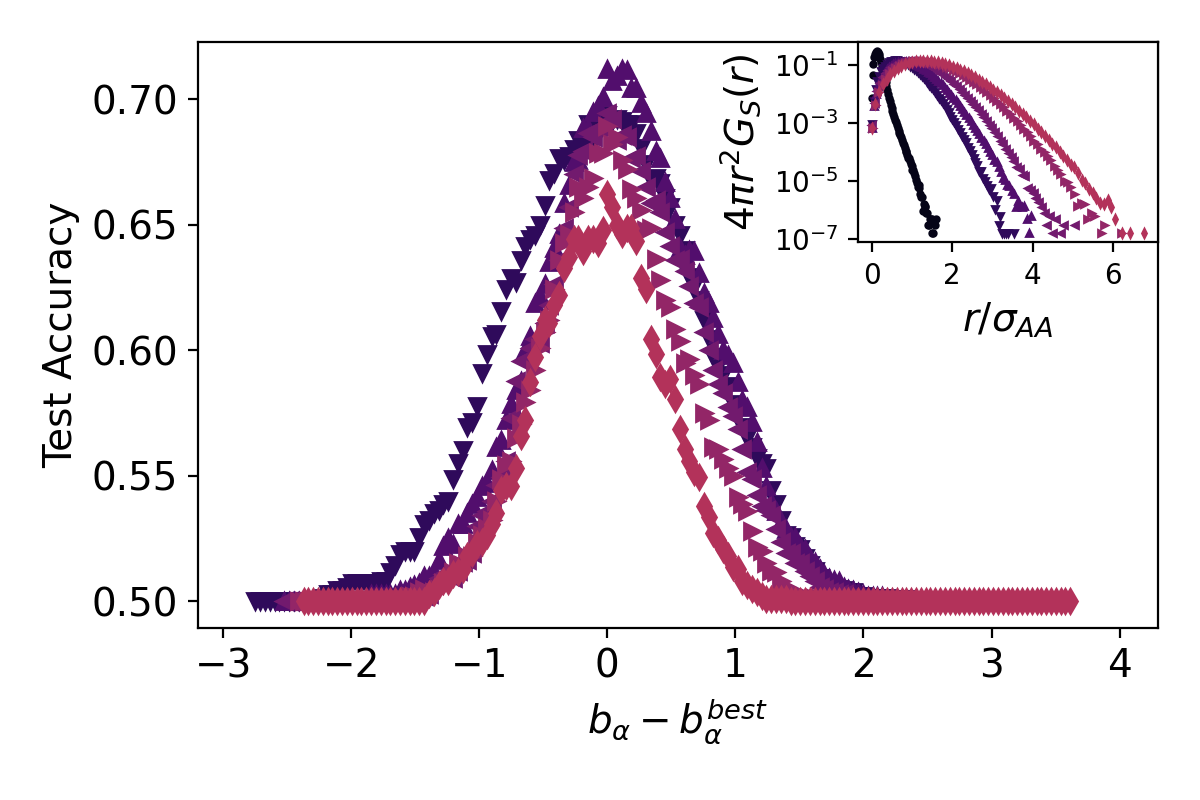}
    \caption{ \textbf{A transfer learning approach connects extreme diffusive events above $T_0$ with activated dynamics below $T_0$}.  The main figure shows the test accuracy  of linear SVMs, trained on extreme diffusive samples at $T=1.0,1.2,1.4,1.8,2.0$ (dark blue to light red), as applied to a test set of activated dynamics at $T=0.47$ as a function of the classifier's bias relative to the optimal bias for that choice of temperature. The inset show the self part of the Van-Hove correlation function for large particles at a time scale of 10 LJ time units. (Black dots are for T=0.45.)}
    \label{fig:transferLearning}
\end{figure}

We believe it is noteworthy that at such high temperatures, \emph{any} structural features predictive of dynamics can be found. We find that even a softness classifier -- i.e., a classifier trained on activated dynamics -- has some ability to classify diffusive events in the fluid phase: as shown in the inset, the accuracy on the high-$T$ training sets is almost as good as the classifiers trained at those temperatures. The optimal bias that needs to be chosen is quite different, but the \emph{direction} in feature space learned is quite similar. This finding encourages us to more explicitly frame a transfer learning task from the high-temperature to the low-temperature regime. Concretely, we apply the fluid-phase classifiers -- trained at temperatures ranging from $T=1$ to $T=2$ -- to labeled data from $T=0.47$. As shown in Fig.~\ref{fig:transferLearning}, even though we have trained on data well above $T_0$ we find that our classifiers maintain substantial classification accuracy. Again, the optimal bias varies strongly with training and testing temperature, but the \emph{direction} in feature space is extremely highly correlated. 

To highlight how surprising this is, in the inset we show the self part of the van Hove function characterizing single-particle displacements, choosing as a time scale the same window we used for $p_{hop}$. At high temperatures this distribution is essentially Gaussian and involves a substantial numbers of particles moving \emph{many} times their own size; at low temperatures this distribution is non-trivial and has an exponential tail corresponding to hopping motions whose size is less than a single particle diameter.

\section{IV. Interpretability of fluidity and softness}

\begin{figure}[t!]
    \centering
    \includegraphics[width =1.0\linewidth]{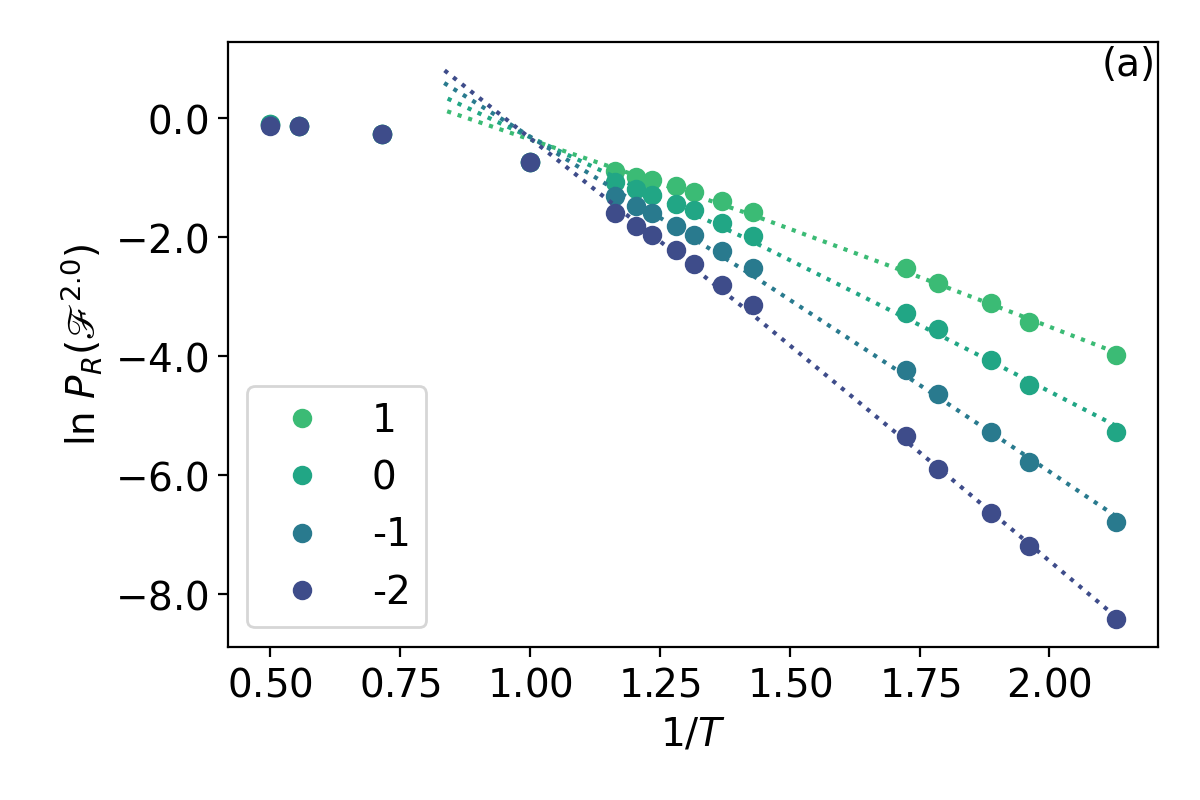}
    \includegraphics[width =1.0\linewidth]{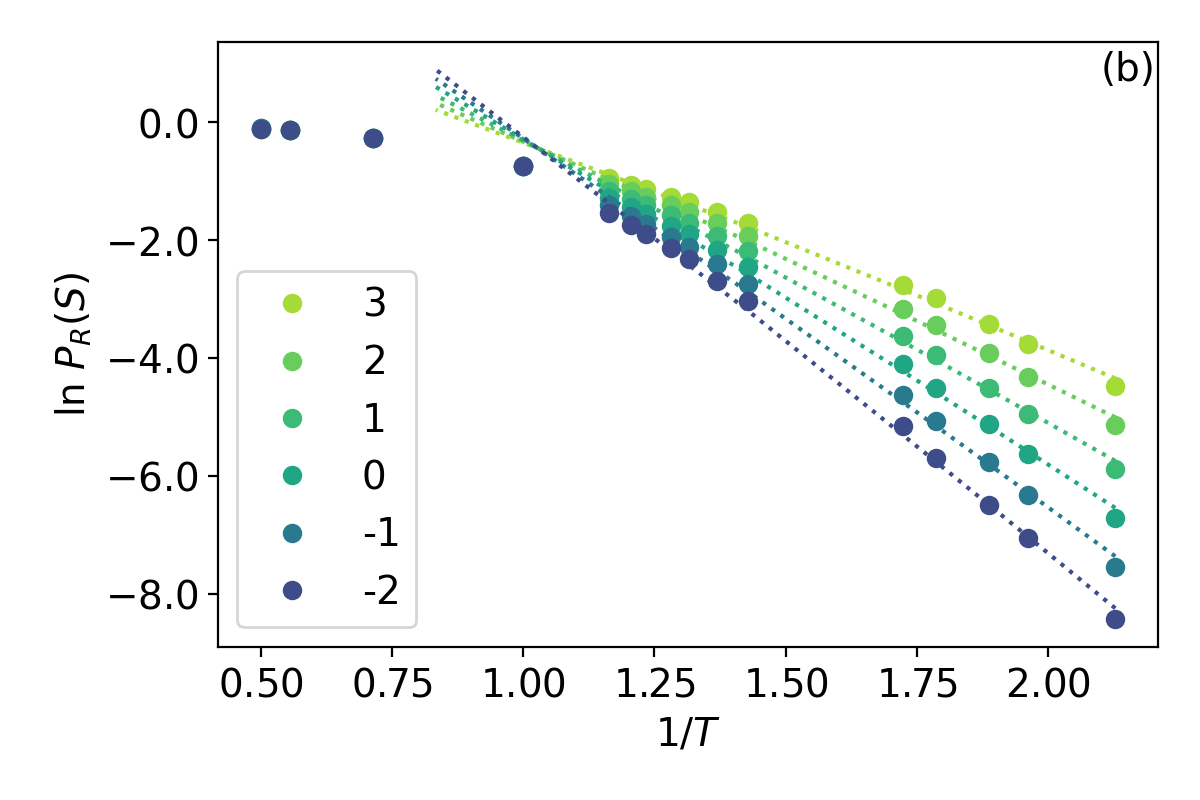}
    \caption{\textbf{The probability of rearrangement conditioned on fluidity reveals energy barriers below the onset temperature}. (a) The log probability of rearrangement conditioned on $\mathcal{F}^{2.0}$ vs inverse temperature. Point colors correspond to different bins of fluidity, as indicated in the legend. Part (b) shows the same features for rearrangements conditioned on $S$. In all cases, dotted lines are Kramers-form fits.}
    \label{fig:physicalInterpretationOfClassifiers}
\end{figure}

Using $T>T_0$ classifiers we are able to obtain reasonable accuracy on training sets (which are by definition constructed from atypical particles at the various training temperatures), but remarkably we find that fluidity has the same kind of physical interpretability as softness. We define a rearrangement as a particle having an instantaneous value of $p_{hop}>p_c$, and fit the probability of rearranging, $P_R$, to a Kramers form \cite{hanggi1990reaction}: $P_R=\frac{1}{T}\exp\left( \Sigma(\mathcal{F}^T)\right) \exp \left(- \Delta E(\mathcal{F}^T)/T\right)$. Just as for softness, we show in Fig.~\ref{fig:physicalInterpretationOfClassifiers} that fluidity partitions the overall system dynamics into a collection of barrier-hopping processes characterized by an energy barrier scale ($\Delta E$) and an entropic contribution ($\Sigma$).  We also find that our prediction of the onset temperature itself -- whether from the intersection of the Kramers form fits or more qualitatively from where the data collapses -- is the same across our softness and fluidity classifiers, suggesting that a consistent physical interpretation is being learned.

The identification of a scalar value -- fluidity -- that encodes the energy barrier characterizing an activated process  by training a classifier on diffusive events is striking. Given the cross-over nature of the onset temperature, perhaps this qualitative result could have been expected for training temperatures close to $T_0$, but it holds even when training far above $T_0$, as shown in Fig.~\ref{fig:physicalInterpretationOfClassifiers}a. How do the energy barriers learned by these classifiers compare to the energy barriers learned by classifiers trained on supercoooled data, i.e., to those from softness? A direct answer to this question is complicated by two aspects of the training and testing procedure. 

\begin{figure}[ht]
    \centering
    \includegraphics[width =1\linewidth]{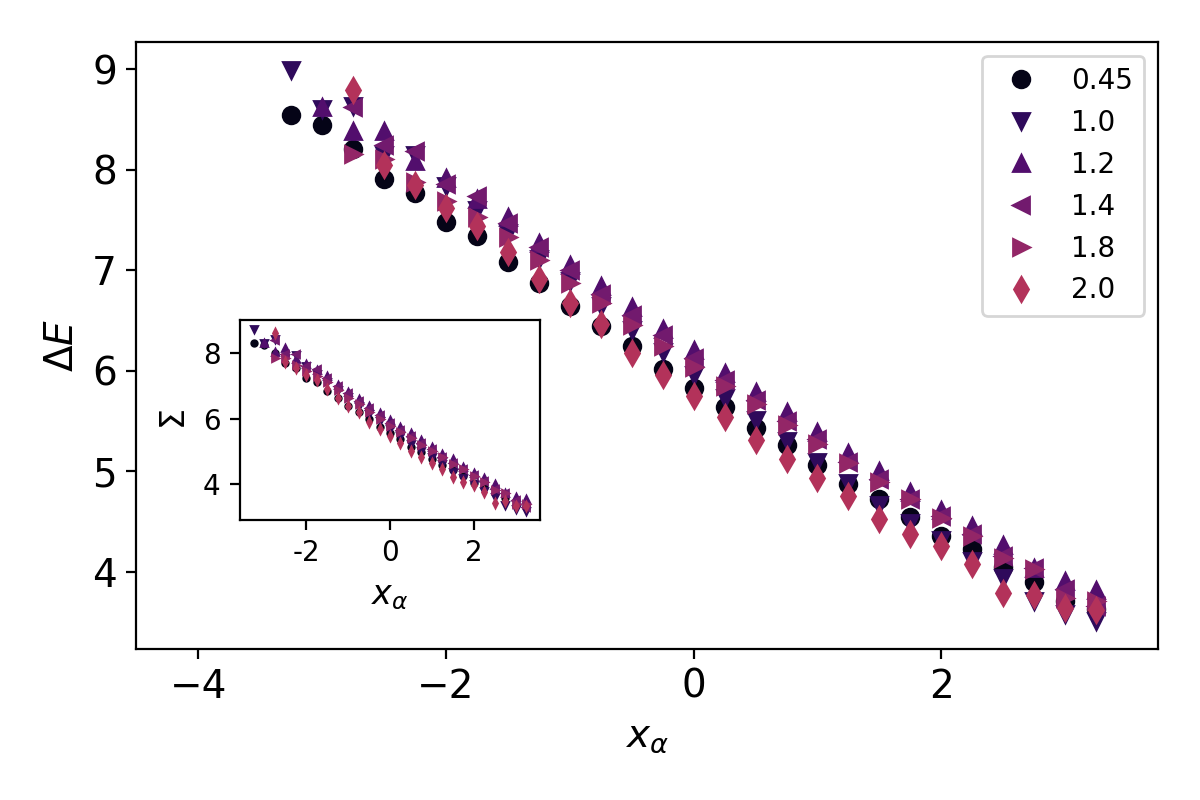}
    \caption{\textbf{Collapse of inferred landscape features from different
training temperatures.} The energy barrier as a function of distance to optimized classifier, $\Delta E$ vs. $x_\alpha$, as inferred from Kramers fits to $P_R(S)$ shows little variation across a wide range of classifier training temperatures. The inset showing the entropic contribution similarly collapses in this representation.}
    \label{fig:barrierVsClassifierDirection}
\end{figure}

The first is that there is no reason to think that the hyperplane \emph{bias} should be held constant when moving from one task to another. This is implicit in the relatively large shifts in bias needed in the inset of  Fig.~\ref{fig:classificationAccuracy} and in the test accuracy for sub-optimal choices of bias in Fig.~\ref{fig:transferLearning}. The second issue relates to the fact that  we study systems across such a wide temperature range that the distribution of the structural features changes substantially (a similar issue arose in the context of applying classifiers to systems at different \emph{densities} \cite{tah2022fragility}). To account for these, we compare the physical interpretations of the different classifiers by defining 
$x_\alpha = \left(\vec{w}_\alpha\cdot \vec{F} -b_\alpha^{best}\right)/ \sigma_\alpha$.
That is, we  adjust the bias to the optimal value when the classifier is applied to a common ($T=0.47$)  training set, and rescale the feature vector by the standard deviation of the distribution of fluidity (or softness) at the training temperature. With this choice, in Fig.~\ref{fig:barrierVsClassifierDirection} we show that the learned aspects of the landscape associated with particle structure -- including both the energy barrier and entropic contribution -- are almost \emph{identical}. We note that fitting the data only in the regime unambiguously below the onset temperature -- i.e., the points for which $T<0.8$ -- does not qualitatively change these results. We speculate that there may be some correlation between training at higher temperatures and a hint of a slight curvature in the data, but do not yet have sufficient data to confirm this.


Given these results, Table \ref{tab:trainingTemperaturebandsigma_phop} reports several values that contribute to the formation of the results: the bias (i.e., the bias that achieves the highest accuracy during training), standard deviation of fluidity and softness, and the optimal choice of bias when the classifier is applied to a test set at $T=0.47$. We also report, as a simple measure of the similarity of the classifiers, the dot product between the normal vector describing each classifier and that of the softness classifier.

\begin{table}
    \centering
    \begin{tabular}{c|c|c|c|c}
         \hline\hline
         $T$ & $b$ & $\sigma_T$ & $b^{best}$ & $\hat{w}_{S} \cdot \hat{w}_{\alpha}$\\
     0.45  & 0.0096  & 1.1188 & 0.55710 & 1.0000\\
     1.0  & -0.0056  & 0.7653 & -0.2385 & 0.8246\\
     1.2  & -0.0171  & 0.7028 & -0.4644 & 0.8071\\
     1.4  & -0.0099  & 0.6760 & -0.4644 & 0.6643\\
     1.8  & -0.0262  & 0.6286 & -0.5774 & 0.7104\\
     2.0  & -0.0341  & 0.6286 & -0.6151 & 0.7198\\
    \end{tabular}
    \caption{\textbf{Table of the optimal bias, b, of the hyperplane during training; the standard deviation of fluidity (or softness for T=0.45) at the training temperatures; and the optimal bias, $b^{best}$, that maximizes test accuracy at T=0.47. The final column displays the projection of our classifiers onto the softness classifier.}}
    \label{tab:trainingTemperaturebandsigma_phop}
\end{table} 

\section*{V. Discussion}

Taken together, our work establishes a surprising connection between the structural features that control activated events at low temperatures and those apparently responsible for the tails of the distribution of diffusive events above the onset temperature. Although many approaches have considered the link between local structural arrangements and dynamical arrest in the supercooled regime \cite{royall2015role}, much of this knowledge is set aside when studying the liquid phase. Our finding that structure is relevant even above the temperature of liquid-gas critical point (roughly $T=1.2$ in this model \cite{testard2014intermittent}) suggests that further pursuing this avenue of research may prove fruitful. The connections between structure and dynamics across temperatures that we find may be a consequence of the only modestly growing structural length scales over the temperature range studied, but we again emphasize that the qualitative character of the dynamics changes significantly over these same temperatures.

A natural hypothesis might be that our classification accuracy stems from an ability to identify fluid phase particles that do not diffuse very much: perhaps we are identifying rare particles that consistently sample a similar, high-barrier part of the energy landscape, and are not truly distinguishing both immobile and highly-mobile particles? We show in the Appendix that this hypothesis fails, and that using both tails of the diffusive-motion distribution is crucial to our results. We comment that our main finding --  that one can take a classifier built on fluid-phase data without barrier-hopping dynamics, apply it to data in a dynamically heterogeneous phase, and infer the existence of energy barriers there -- is reminiscent of the results reported in Ref.~\cite{tah2021quantifying}. That work considered a biologically-inspired model with highly unusual glassy dynamics \cite{sussman2018anomalous,li2021softness,pinto2022hierarchical} meant to mimic the behavior of dense cellular materials. There it was speculated that it was the anomalous, sub-Arrhenius behavior of the model  that was responsible for the success of the transfer learning task; the results presented here suggest an alternative explanation may be needed.

Our work highlights what we believe continue to be crucial unanswered questions: \emph{why} do these machine learning methodologies learn simple structural order parameters that correspond to local energy barriers in disordered phases of matter? What aspects of the training lead to this result? And to what extent can we use this result to uncover new, relevant descriptions for the physics of amorphous solids? We note that there is some indication that the specific methodology used here and earlier -- linear SVMs -- may not be crucial to recover this physical interpretation; Ref.~\cite{boattini2021averaging} hinted at a similar result using a GNN-inspired linear-regression-based model. We believe it will be crucial to compare different machine learning techniques as applied to predicting glassy dynamics \cite{alkemade2022comparing,bapst2020unveiling} not only along dimensions of predictive capacity, generalizability, efficiency, and training cost, but \emph{also} in terms of their physical interpretability.

\begin{acknowledgments}

This material is based upon work supported by the National Science Foundation under Grant No.~DMR-2143815. We thank Ilya Nemenman, Sean Ridout, Arabind Swain, Andrea Liu, and Eric Weeks for stimulating discussions. Data used in this paper is available at \cite{obadiyaDataset2022}.
\end{acknowledgments}

\bibliography{FluidityPaper}

\begin{thebibliography}{10}

\bibitem{berthier2011dynamic}
Ludovic Berthier.
\newblock Dynamic heterogeneity in amorphous materials.
\newblock {\em arXiv preprint arXiv:1106.1739}, 2011.

\bibitem{angell1995formation}
C~Austen Angell.
\newblock Formation of glasses from liquids and biopolymers.
\newblock {\em Science}, 267(5206):1924--1935, 1995.

\bibitem{sastry1998signatures}
Srikanth Sastry, Pablo~G Debenedetti, and Frank~H Stillinger.
\newblock Signatures of distinct dynamical regimes in the energy landscape of a
  glass-forming liquid.
\newblock {\em Nature}, 393(6685):554--557, 1998.

\bibitem{debenedetti2001supercooled}
Pablo~G Debenedetti and Frank~H Stillinger.
\newblock Supercooled liquids and the glass transition.
\newblock {\em Nature}, 410(6825):259--267, 2001.

\bibitem{banerjee2017determination}
Atreyee Banerjee, Manoj~Kumar Nandi, Srikanth Sastry, and Sarika
  Maitra~Bhattacharyya.
\newblock Determination of onset temperature from the entropy for fragile to
  strong liquids.
\newblock {\em The Journal of chemical physics}, 147(2):024504, 2017.

\bibitem{folena2020rethinking}
Giampaolo Folena, Silvio Franz, and Federico Ricci-Tersenghi.
\newblock Rethinking mean-field glassy dynamics and its relation with the
  energy landscape: The surprising case of the spherical mixed p-spin model.
\newblock {\em Physical Review X}, 10(3):031045, 2020.

\bibitem{keys2011excitations}
Aaron~S Keys, Lester~O Hedges, Juan~P Garrahan, Sharon~C Glotzer, and David
  Chandler.
\newblock Excitations are localized and relaxation is hierarchical in
  glass-forming liquids.
\newblock {\em Physical Review X}, 1(2):021013, 2011.

\bibitem{ediger2000spatially}
Mark~D Ediger.
\newblock Spatially heterogeneous dynamics in supercooled liquids.
\newblock {\em Annual review of physical chemistry}, 51(1):99--128, 2000.

\bibitem{berthier2011dynamical}
Ludovic Berthier, Giulio Biroli, Jean-Philippe Bouchaud, Luca Cipelletti, and
  Wim van Saarloos.
\newblock {\em Dynamical heterogeneities in glasses, colloids, and granular
  media}, volume 150.
\newblock OUP Oxford, 2011.

\bibitem{ediger1996supercooled}
Mark~D Ediger, C~Austen Angell, and Sidney~R Nagel.
\newblock Supercooled liquids and glasses.
\newblock {\em The journal of physical chemistry}, 100(31):13200--13212, 1996.

\bibitem{pedersen2018phase}
Ulf~R Pedersen, Thomas~B Schr{\o}der, and Jeppe~C Dyre.
\newblock Phase diagram of kob-andersen-type binary lennard-jones mixtures.
\newblock {\em Physical review letters}, 120(16):165501, 2018.

\bibitem{berthier2007structure}
Ludovic Berthier and Robert~L Jack.
\newblock Structure and dynamics of glass formers: Predictability at large
  length scales.
\newblock {\em Physical Review E}, 76(4):041509, 2007.

\bibitem{kawasaki2007correlation}
Takeshi Kawasaki, Takeaki Araki, and Hajime Tanaka.
\newblock Correlation between dynamic heterogeneity and medium-range order in
  two-dimensional glass-forming liquids.
\newblock {\em Physical review letters}, 99(21):215701, 2007.

\bibitem{patrick2008direct}
C~Patrick~Royall, Stephen~R Williams, Takehiro Ohtsuka, and Hajime Tanaka.
\newblock Direct observation of a local structural mechanism for dynamic
  arrest.
\newblock {\em Nature materials}, 7(7):556--561, 2008.

\bibitem{jack2014information}
Robert~L Jack, Andrew~J Dunleavy, and C~Patrick Royall.
\newblock Information-theoretic measurements of coupling between structure and
  dynamics in glass formers.
\newblock {\em Physical review letters}, 113(9):095703, 2014.

\bibitem{manning2011vibrational}
M~Lisa Manning and Andrea~J Liu.
\newblock Vibrational modes identify soft spots in a sheared disordered
  packing.
\newblock {\em Physical Review Letters}, 107(10):108302, 2011.

\bibitem{widmer2008irreversible}
Asaph Widmer-Cooper, Heidi Perry, Peter Harrowell, and David~R Reichman.
\newblock Irreversible reorganization in a supercooled liquid originates from
  localized soft modes.
\newblock {\em Nature Physics}, 4(9):711--715, 2008.

\bibitem{cubuk2015identifying}
Ekin~D Cubuk, Samuel~Stern Schoenholz, Jennifer~M Rieser, Brad~Dean Malone,
  Joerg Rottler, Douglas~J Durian, Efthimios Kaxiras, and Andrea~J Liu.
\newblock Identifying structural flow defects in disordered solids using
  machine-learning methods.
\newblock {\em Physical review letters}, 114(10):108001, 2015.

\bibitem{schoenholz2016structural}
Samuel~S Schoenholz, Ekin~D Cubuk, Daniel~M Sussman, Efthimios Kaxiras, and
  Andrea~J Liu.
\newblock A structural approach to relaxation in glassy liquids.
\newblock {\em Nature Physics}, 12(5):469--471, 2016.

\bibitem{sussman2017disconnecting}
Daniel~M Sussman, Samuel~S Schoenholz, Ekin~D Cubuk, and Andrea~J Liu.
\newblock Disconnecting structure and dynamics in glassy thin films.
\newblock {\em Proceedings of the National Academy of Sciences},
  114(40):10601--10605, 2017.

\bibitem{bapst2020unveiling}
Victor Bapst, Thomas Keck, A~Grabska-Barwi{\'n}ska, Craig Donner, Ekin~Dogus
  Cubuk, Samuel~S Schoenholz, Annette Obika, Alexander~WR Nelson, Trevor Back,
  Demis Hassabis, et~al.
\newblock Unveiling the predictive power of static structure in glassy systems.
\newblock {\em Nature Physics}, 16(4):448--454, 2020.

\bibitem{boattini2020autonomously}
Emanuele Boattini, Susana Mar{\'\i}n-Aguilar, Saheli Mitra, Giuseppe Foffi,
  Frank Smallenburg, and Laura Filion.
\newblock Autonomously revealing hidden local structures in supercooled
  liquids.
\newblock {\em Nature communications}, 11(1):1--9, 2020.

\bibitem{boattini2021averaging}
Emanuele Boattini, Frank Smallenburg, and Laura Filion.
\newblock Averaging local structure to predict the dynamic propensity in
  supercooled liquids.
\newblock {\em Physical Review Letters}, 127(8):088007, 2021.

\bibitem{cubuk2020unifying}
Ekin~D Cubuk, Andrea~J Liu, Efthimios Kaxiras, and Samuel~S Schoenholz.
\newblock Unifying framework for strong and fragile liquids via machine
  learning: a study of liquid silica.
\newblock {\em arXiv preprint arXiv:2008.09681}, 2020.

\bibitem{jung2022predicting}
Gerhard Jung, Giulio Biroli, and Ludovic Berthier.
\newblock Predicting dynamic heterogeneity in glass-forming liquids by
  physics-informed machine learning.
\newblock {\em arXiv preprint arXiv:2210.16623}, 2022.

\bibitem{tah2021quantifying}
Indrajit Tah, Tristan~A Sharp, Andrea~J Liu, and Daniel~M Sussman.
\newblock Quantifying the link between local structure and cellular
  rearrangements using information in models of biological tissues.
\newblock {\em Soft Matter}, 2021.

\bibitem{kob1995testing}
Walter Kob and Hans~C Andersen.
\newblock Testing mode-coupling theory for a supercooled binary lennard-jones
  mixture i: The van hove correlation function.
\newblock {\em Physical Review E}, 51(5):4626, 1995.

\bibitem{frenkel2023understanding}
Daan Frenkel and Berend Smit.
\newblock {\em Understanding molecular simulation: from algorithms to
  applications}.
\newblock Elsevier, 2023.

\bibitem{anderson2020hoomd}
Joshua~A Anderson, Jens Glaser, and Sharon~C Glotzer.
\newblock Hoomd-blue: A python package for high-performance molecular dynamics
  and hard particle monte carlo simulations.
\newblock {\em Computational Materials Science}, 173:109363, 2020.

\bibitem{behler2007generalized}
J{\"o}rg Behler and Michele Parrinello.
\newblock Generalized neural-network representation of high-dimensional
  potential-energy surfaces.
\newblock {\em Physical review letters}, 98(14):146401, 2007.

\bibitem{hsu2003practical}
Chih-Wei Hsu, Chih-Chung Chang, Chih-Jen Lin, et~al.
\newblock A practical guide to support vector classification, 2003.

\bibitem{candelier2009building}
Raphael Candelier, Olivier Dauchot, and Giulio Biroli.
\newblock Building blocks of dynamical heterogeneities in dense granular media.
\newblock {\em Physical review letters}, 102(8):088001, 2009.

\bibitem{pedregosa2011scikit}
Fabian Pedregosa, Ga{\"e}l Varoquaux, Alexandre Gramfort, Vincent Michel,
  Bertrand Thirion, Olivier Grisel, Mathieu Blondel, Peter Prettenhofer, Ron
  Weiss, Vincent Dubourg, et~al.
\newblock Scikit-learn: Machine learning in python.
\newblock {\em the Journal of machine Learning research}, 12:2825--2830, 2011.

\bibitem{bocquet2009kinetic}
Lyd{\'e}ric Bocquet, Annie Colin, and Armand Ajdari.
\newblock Kinetic theory of plastic flow in soft glassy materials.
\newblock {\em Physical review letters}, 103(3):036001, 2009.

\bibitem{hanggi1990reaction}
Peter H{\"a}nggi, Peter Talkner, and Michal Borkovec.
\newblock Reaction-rate theory: fifty years after kramers.
\newblock {\em Reviews of modern physics}, 62(2):251, 1990.

\bibitem{tah2022fragility}
Indrajit Tah, Sean~A Ridout, and Andrea~J Liu.
\newblock Fragility in glassy liquids: A structural approach based on machine
  learning.
\newblock {\em arXiv preprint arXiv:2205.07187}, 2022.

\bibitem{royall2015role}
C~Patrick Royall and Stephen~R Williams.
\newblock The role of local structure in dynamical arrest.
\newblock {\em Physics Reports}, 560:1--75, 2015.

\bibitem{testard2014intermittent}
Vincent Testard, Ludovic Berthier, and Walter Kob.
\newblock Intermittent dynamics and logarithmic domain growth during the
  spinodal decomposition of a glass-forming liquid.
\newblock {\em The Journal of chemical physics}, 140(16):164502, 2014.

\bibitem{sussman2018anomalous}
Daniel~M Sussman, M~Paoluzzi, M~Cristina Marchetti, and M~Lisa Manning.
\newblock Anomalous glassy dynamics in simple models of dense biological
  tissue.
\newblock {\em EPL (Europhysics Letters)}, 121(3):36001, 2018.

\bibitem{li2021softness}
Yan-Wei Li, Leon Loh~Yeong Wei, Matteo Paoluzzi, and Massimo~Pica Ciamarra.
\newblock Softness, anomalous dynamics, and fractal-like energy landscape in
  model cell tissues.
\newblock {\em Physical Review E}, 103(2):022607, 2021.

\bibitem{pinto2022hierarchical}
Diogo~EP Pinto, Daniel~M Sussman, Margarida M~Telo da~Gama, and Nuno~AM
  Ara{\'u}jo.
\newblock Hierarchical structure of the energy landscape in the voronoi model
  of dense tissue.
\newblock {\em Physical Review Research}, 4(2):023187, 2022.

\bibitem{alkemade2022comparing}
Rinske~M Alkemade, Emanuele Boattini, Laura Filion, and Frank Smallenburg.
\newblock Comparing machine learning techniques for predicting glassy dynamics.
\newblock {\em The Journal of Chemical Physics}, 156(20):204503, 2022.

\bibitem{obadiyaDataset2022}
Tomilola~M Obadiya and Daniel~M Sussman.
\newblock Dataset for ``does fluid structure encode predictions of glassy
  dynamics?''.
\newblock \url{http://dx.doi.org/110.5281/zenodo.7469766}.

\end{thebibliography}

\appendix
\section{APPENDIX}
\label{appendix}

\subsection{Fluidity Distribution}\label{sec:fluidityDistribution}
\label{sec:FluidDist}

\begin{figure}[ht]
    \centering
    \includegraphics[width =1.0\linewidth]{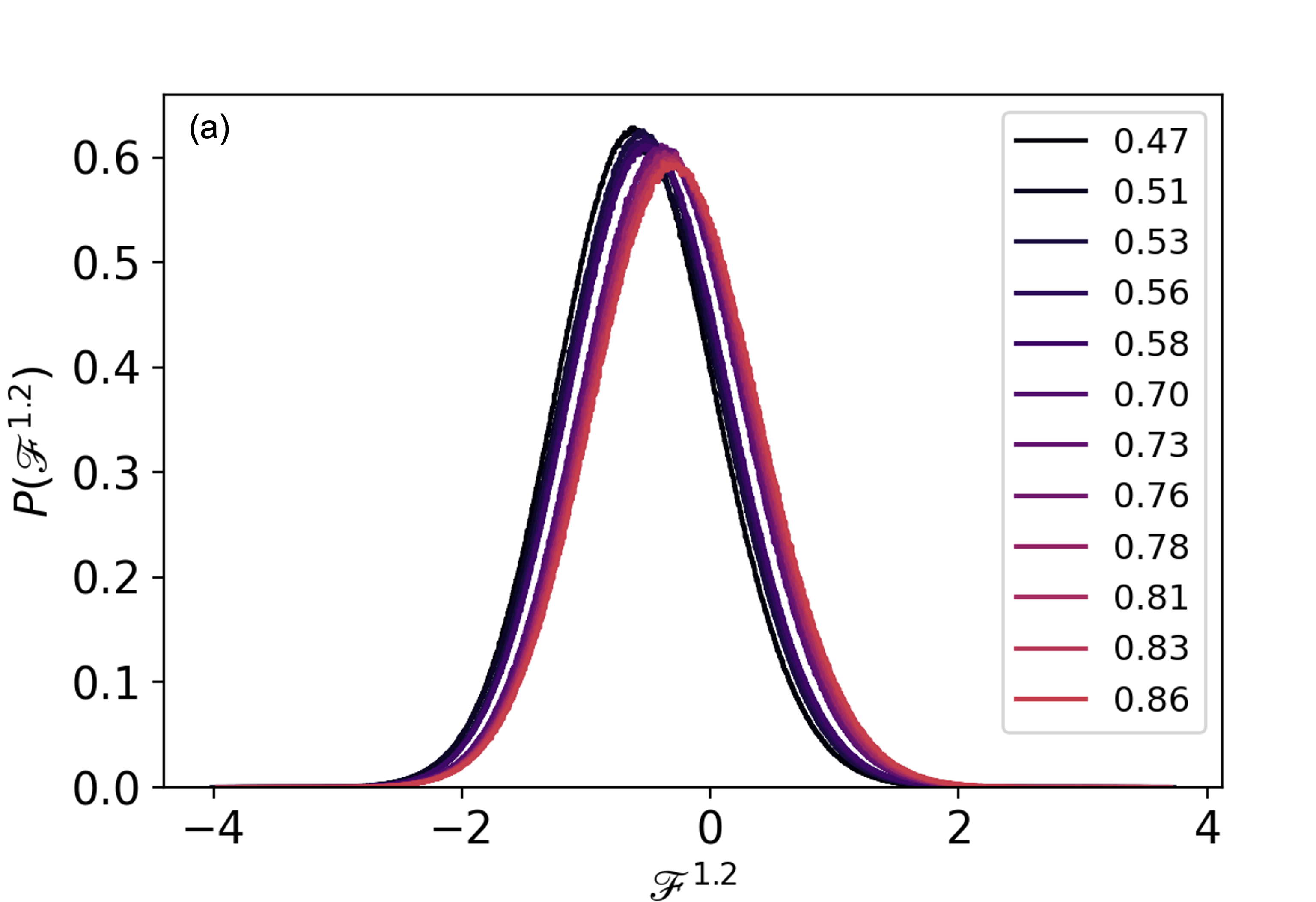}
    \includegraphics[width =1.0\linewidth]{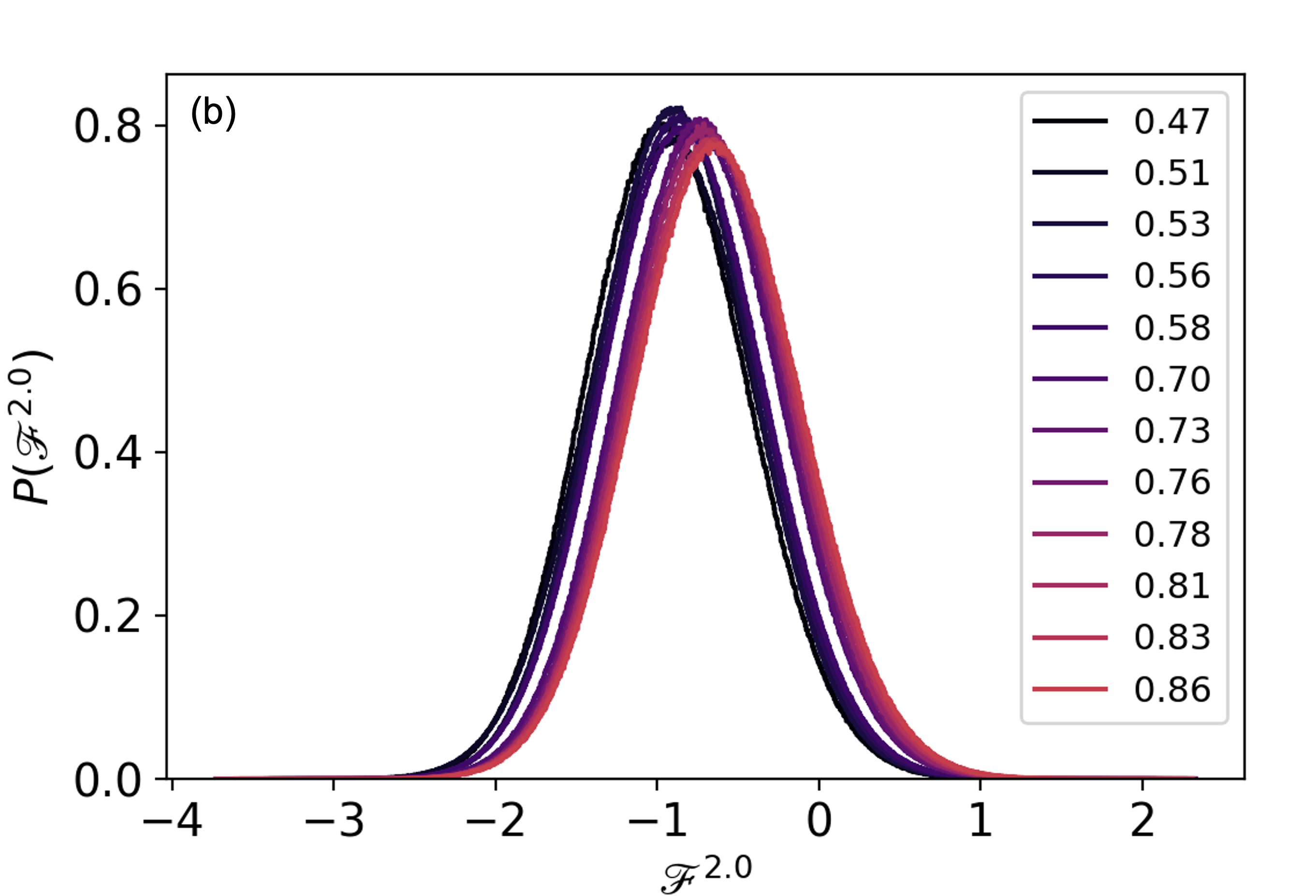}
    \caption{ \textbf{The distribution of fluidity  at different test temperatures for two considered training temperatures.} Training at both $T=1.2$  (a) and $T=2.0$ (b) the distribution of fluidity is approximately Gaussian. The mean is a monotonic function of the test temperature.}
    \label{fig:fluiditydistributionforT1.2_2}
\end{figure}

Similar to the distribution of softness shown in Ref.~\cite{schoenholz2016structural}, the distributions of fluidity, when measured at different test temperatures, remains approximately Gaussian. This is shown in Fig.~\ref{fig:fluiditydistributionforT1.2_2} for two training temperatures ($T= 1.2$ and $T=2.0$). The mean of the distribution behaves monotonically as the test temperature changes.

\subsection{Determination of rearrangement threshold for fluidity}
\label{sec:DoRTF}
\begin{figure}[t]
\centering
    \includegraphics[width =1.0\linewidth]{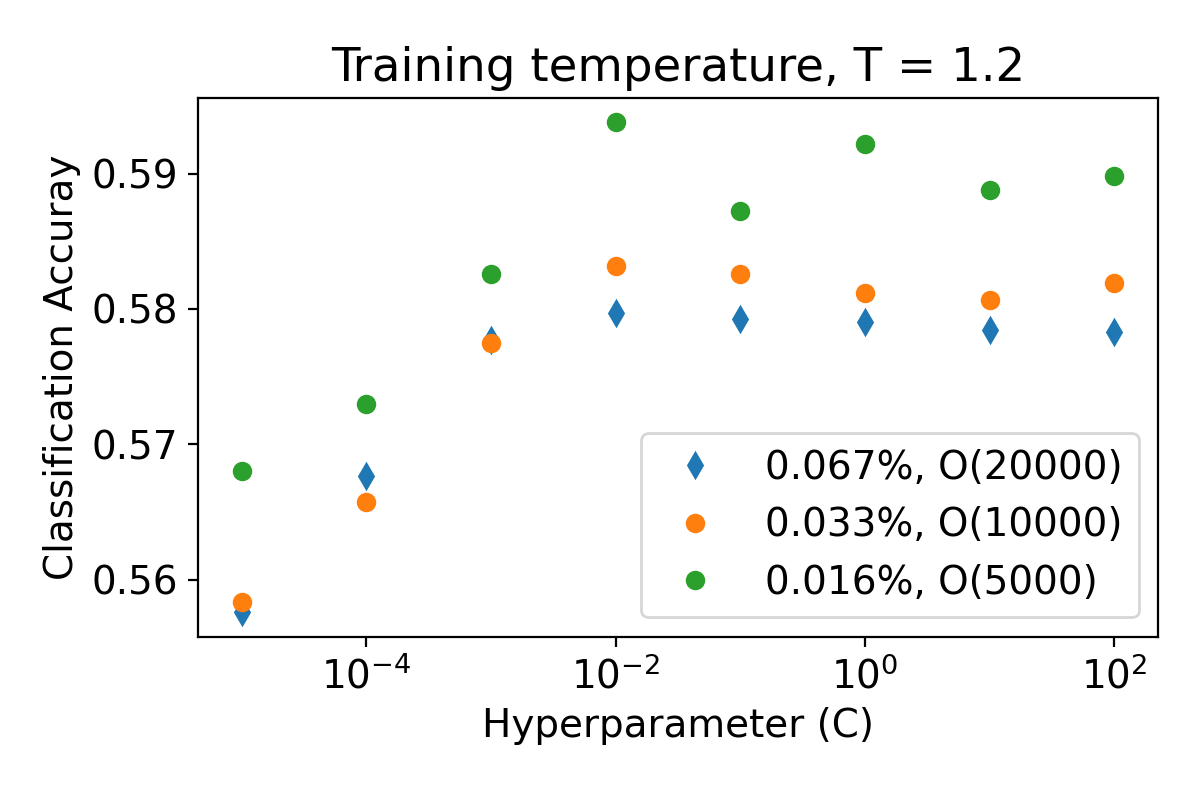}
    \caption{\textbf{Classification accuracy above the onset temperature as a function of misclassification penalty (C) using different thresholds for defining extreme events.} Three different choices for the fraction of the distribution of events to use in the training set give broadly consistent accuracies on the test set, with no systematic changes in the optimal hyperparameter.}
    \label{fig:thresholdplot}
\end{figure}

\begin{figure}[ht]
\centering
    \includegraphics[width =1.0\linewidth]{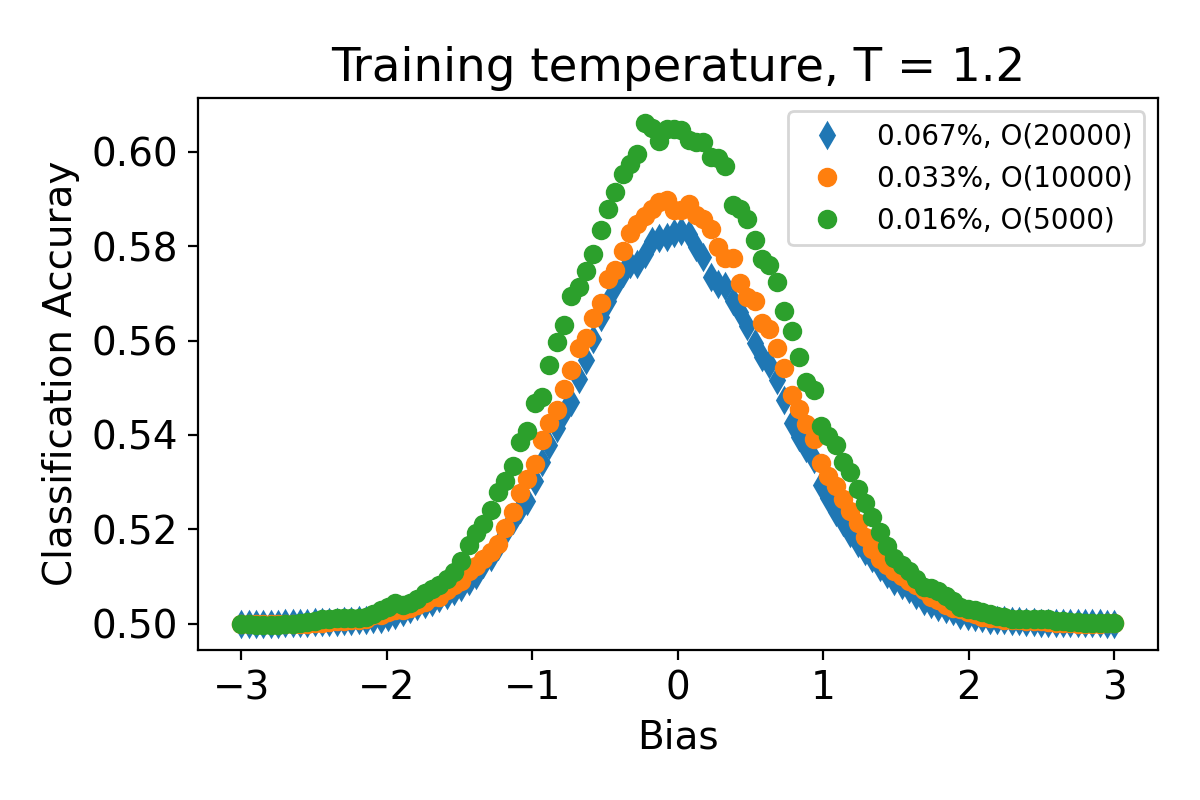}
    \caption{\textbf{Classification accuracy above the onset temperature as a function of bias at fixed training hyperparameter}. Using a fixed misclassification hyperparameter of $C=10^{-2}$, we again see that the accuracy is only weakly dependent on the chosen event thresholds.}
    \label{fig:thresholdplot_bias}
\end{figure}

The dependence of softness on various choices for training set thresholds was explored in Ref.~\cite{schoenholz2016structural}, and our choices were consistent with widely employed values. In the case of fluidity, as noted above, we selected threshold values to have comparably sized training sets given the length of our simulations (i.e., to generate of order $10^4$ elements of a balanced training set). We first emphasize that the qualitative outcome of our analysis is not crucially dependent on the choice of threshold. In Fig.~\ref{fig:thresholdplot} and \ref{fig:thresholdplot_bias} we show (at a fixed representative temperature above the onset temperature, $T=1.2$) that our classification accuracy on the test set is only weakly dependent on the precise choice of $p_l$ and $p_u$: sensibly, more extreme events are slightly easier to classify, but we are far enough into the tail of rare events that the effect is a small one. 

\begin{figure}[ht]
\centering
    \includegraphics[width =1.0\linewidth]{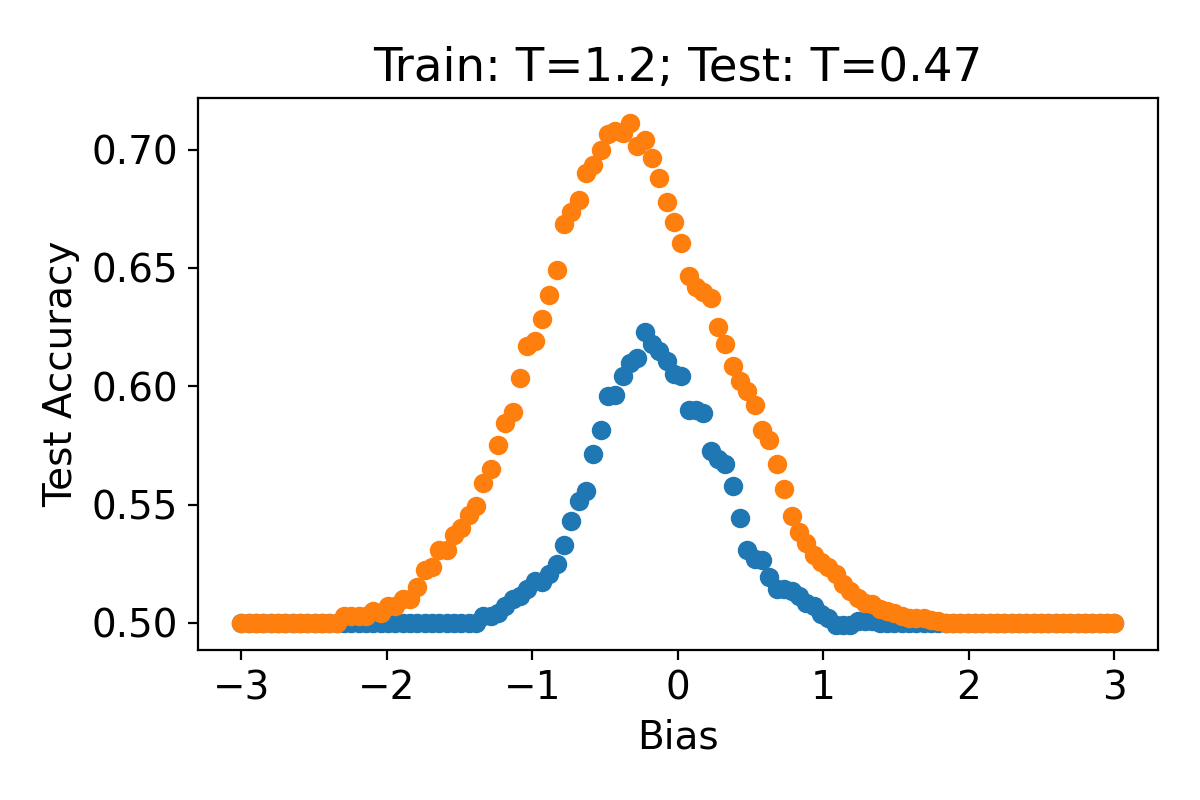}
    \caption{\textbf{Test accuracy against bias of a ``low mobility vs random particle'' classifier}. The plot shows that the test accuracy of the classifier (blue, lower points) is lower than the classifier built using both extreme ends of the $p_{hop}$ distribution (orange, upper points).}
    \label{fig:extremecheck_bias}
\end{figure}

We note, however, that our ability to classify and to successfully transfer learn \emph{is} dependent on the selection of appropriate structural examples that correspond to both extreme ends of the $p_{hop}$ distribution. As discussed in the main text, one natural hypothesis is that local structures corresponding to the lower tail of events capture sufficiently deep minima that influence even particles in the fluid phase, and that our ability to classify stems entirely from these rare, relatively immobile particles. To test this, we construct a classifier using local structures from the lower tail of the $p_{hop}$ distribution as our non-rearranging particles and construct the other class from \emph{randomly selected} particles with $p_{hop}>p_l$. We train such a classifier at $T=1.2$ and evaluate it on the test set of $T=0.47$ (as in Fig. 2 of the main text). We show in Fig.~\ref{fig:extremecheck_bias} that the accuracy of this ``immobile vs. randomly selected'' classifier is substantially lower than the accuracy of the ``immobile vs highly mobile'' fluidity classifier. 

\subsection{Probability of rearrangements as a function of $x_{\alpha}$} \label{sec:PofR}

\begin{figure}[htp]
    \centering
    \includegraphics[width =2\linewidth]{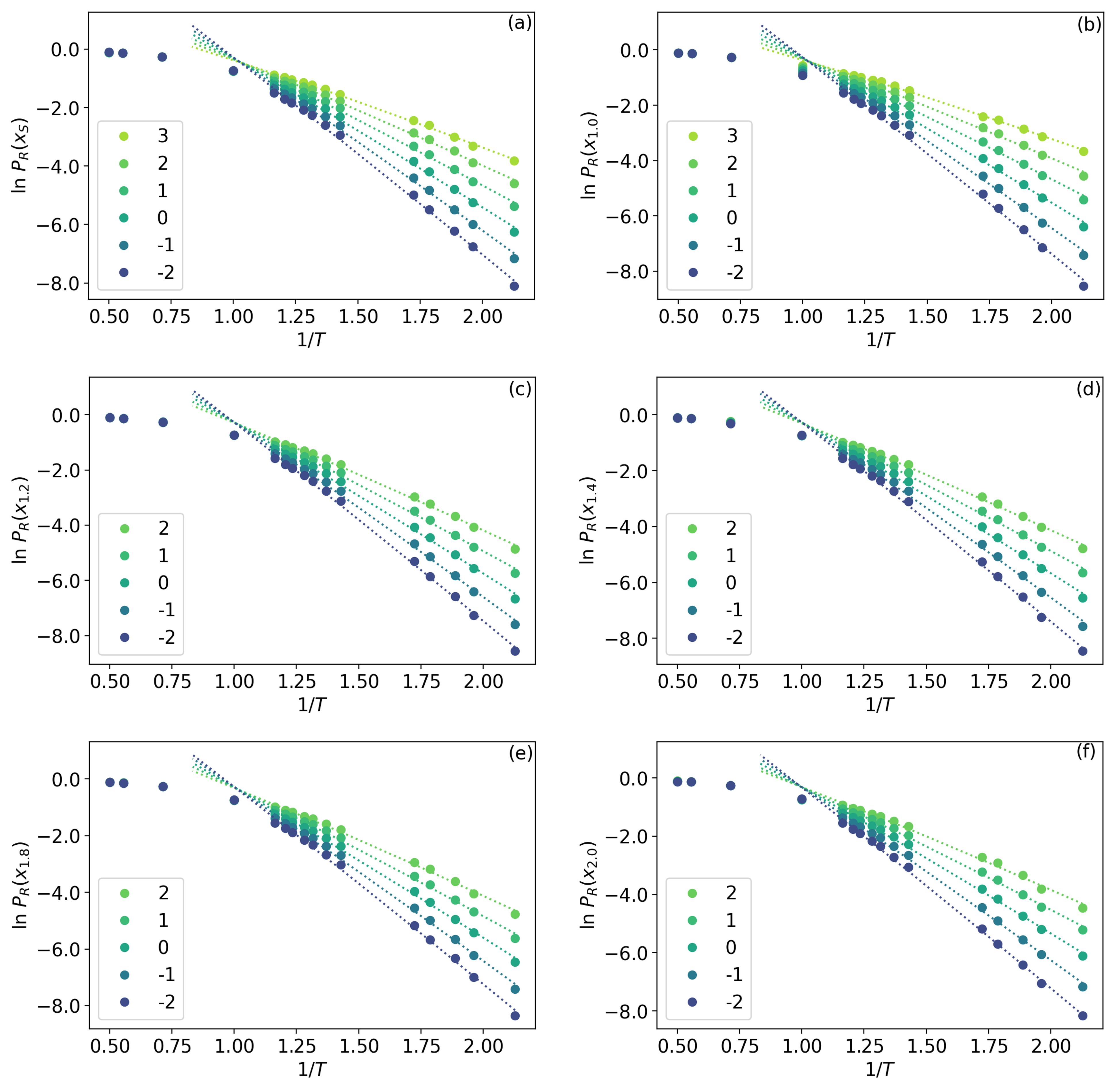}
    \caption{ \textbf{The probability of rearrangement conditioned on $x_{\alpha}$ against 1/T.} Dotted lines are fits to the Kramers form for different bins of $x_\alpha$. }
    \label{fig:kramers_form_fitting_phop}
\end{figure}

In this section, we show  data for the probability of rearrangement as a function of $x_{\alpha}$ against 1/T for different training temperatures -- this data leads directly to the results reported in Fig.~\ref{fig:barrierVsClassifierDirection} of the main text. The probability of rearrangement conditioned on $x_{\alpha}$ is the fraction of local structure with value $x_{\alpha}$ that has a corresponding $p_{hop}$ value greater than $p_c$, where $p_c$ is the rearrangement cutoff used at low T. Figure \ref{fig:kramers_form_fitting_phop} -- mirroring Fig.~\ref{fig:physicalInterpretationOfClassifiers} -- shows the probability of rearrangement as a function of $1/T$ for the softness and the fluidity classifiers. In Fig.~\ref{fig:barrierVsClassifierDirection}, we reported the inferred energy barriers and ``entropic barriers'' associated with each of these fits.

\end{document}